\documentclass[aps,prd,eqsecnum,preprint,tightenlines,nofootinbib]{revtex4-2}
\usepackage{graphicx,latexsym}

\def\bea{\begin{eqnarray}}
\def\eea{\end{eqnarray}}
\def\be{\begin{equation}}
\def\ee{\end{equation}}

\begin{document}

\title{Gravitational soliton solutions \\
to self-coupled Klein-Gordon and Schr\"odinger equations
}
\author{D.A.~Taylor}
\affiliation{Department of Civil and Environmental Engineering, 
Idaho State University, Pocatello, Idaho 83209 USA}
\author{S.S.~Chabysheva}
\affiliation{Department of Physics, University of Idaho, Moscow, Idaho 83844 USA}
\author{J.R.~Hiller}
\affiliation{Department of Physics, University of Idaho, Moscow, Idaho 83844 USA}
\affiliation{Department of Physics and Astronomy,
University of Minnesota-Duluth,
Duluth, Minnesota 55812 USA}

\date{\today}

\begin{abstract}

We use the Klein--Gordon equation in a curved spacetime to construct the
relativistic analog of the Schr\"odinger--Newton problem, where a
scalar particle lives in a gravitational potential well generated by
its own probability distribution.  A static, spherically symmetric
metric is computed from the field equations of general relativity,
both directly and as modeled by a perfect-fluid assumption that uses
the Tolman--Oppenheimer--Volkov equation for hydrostatic equilibrium
of the mass density.  The latter is appropriate for a Hartree
approximation to the many-body problem of a bosonic star.
Simultaneous self-consistent solution of the Klein--Gordon equation in
this curved spacetime then yields solitons with a range of radial
excitations.  We compare results with the nonrelativistic case.

\end{abstract}


\maketitle

\section{Introduction}  \label{sec:Introduction}

If the probability distribution $|\Psi|^2$ of a particle with mass $m$ is interpreted
as a mass distribution $m|\Psi|^2$, a gravitational self-coupling 
can be considered.  This was first applied to bosonic stars
by Ruffini and Bonazzola~\cite{Ruffini} and then later
considered as a mechanism for wave-function collapse~\cite{Diosi,Penrose}.
As formulated in a nonrelativistic context, this is a coupling
between the Schr\"odinger equation and Newtonian gravity.
This Schr\"odinger--Newton problem has been studied extensively
with numerical techniques~\cite{Moroz,Bernstein,Tod,HarrisonMorozTod,Harrison}.
It can be viewed as arising from a semi-classical formulation
of gravity~\cite{Bahrami}, where matter is quantized but
gravity is not\footnote{In this context, the effect of wave-function collapse 
appears to be inconsistent with causality \protect\cite{Bahrami}.}
and where a nonrelativistic limit is taken~\cite{Giulini}.

We instead directly consider the relativistic problem
of a scalar field bound in a spacetime curved
by the probability distribution for its own mass.\footnote{Self
gravitation of a Dirac field can also be considered.  For a 
recent discussion, see \cite{Kain}.}  The matter
field contributes to the stress-energy tensor that acts
as a source term for the general relativistic (GR) equations that determine
the metric~\cite{Carroll,ExactSolns}.  The reduction of the 
GR equations for a spherically symmetric spacetime
is known~\cite{Giulini}.  The matter obeys the Klein-Gordon (KG)
equation in curved spacetime or, in the nonrelativistic case, the Schr\"odinger
equation.  The two sets of equations, GR and KG, must be solved
simultaneously.

For the hypothesized bosonic star, the matter equation
can be viewed as a Hartree approximation to the many-body 
problem.  This can be combined with the assumption of
a perfect fluid in hydrostatic equilibrium, which leads to
the Tolman--Oppenheimer--Volkov (TOV) equation
for the pressure~\cite{Tolman,Oppenheimer,Carroll,ExactSolns}.
The metric is then that of a perfect fluid with a
pressure determined by the TOV equation.  This equation
and the KG equation are again solved self consistently.

The GR/KG system of equations can also be treated in approximation
via expansions in $v/c$ and $\hbar$.  Relativistic corrections 
to the Schr\"odinger--Newton problem, up to first post-Newtonian 
order, have been considered by Brizuela and Duran-Cabac\'es~\cite{Brizuela}. 
Giulini and Gro{\ss}ardt~\cite{Giulini} consider a WKB-type 
expansion.  However, such approximations are not necessary because
the original system of equations can be solved numerically.

For the numerical calculation, we apply a finite-difference
approximation to the radial part of the KG equation,
which then becomes a matrix eigenvalue problem.  The
equations for the metric are solved on the same discrete
grid by a Runge-Kutta algorithm with an error term 
consistent with the finite-difference approximation.
The two sets are solved self consistently by iteration
from an initial guess.

We first consider the 
nonrelativistic Schr\"odinger--Newton problem, in 
Sec.~\ref{sec:SchrodingerNewton}.  This provides a
basis for comparison in considering the Einstein--Klein--Gordon
soliton in Sec.~\ref{sec:KleinGordon}, which recovers the 
nonrelativistic results in the correct limit. In this section
we treat both the perfect-fluid model and the direct solution
of the GR equations.  Throughout, we use units where $\hbar$ 
and $c$ are 1 but keep Newton's gravitational constant $G$ explicit.
We do limit our discussion to spherically symmetric solutions; however,
axially symmetric solutions have been considered for the 
nonrelativistic case~\cite{Harrison,Guzman}.

\section{Schr\"odinger--Newton solitons}  \label{sec:SchrodingerNewton}

We assume spherical symmetry and therefore consider only
S states for the Schr\"odinger equation.  The angular 
part of the wave function then being trivial, we focus
on the radial equation for the reduced wave function $u(r)=rR(r)$,
\be \label{eq:SchrEqn}
-\frac{1}{2m}\frac{d^2u}{dr^2}+V(r)u(r)=Eu(r).
\ee
The self-coupling is through the potential $V$, which is computed
as the gravitational potential generated by a mass
distribution $\rho(r)=m|\psi|^2$ with $\psi=u(r)/r\sqrt{4\pi}$.
This assumes a normalization of $u$ as $\int_0^\infty |u|^2 dr=1$.

Inside the mass distribution, the gravitational field magnitude $F_G$
is determined by Gauss' law to be
\be
F_G(r)=\frac{4\pi G}{r^2}\int_0^r \rho(r')r^{\prime 2}dr'.
\ee
With the potential chosen to be zero at infinity, the
potential function is 
\be
V(r)=\int_\infty^r mF_G(r') dr'
    =4\pi Gm\int_\infty^r \frac{dr'}{r^{\prime 2}}
                       \int_0^{r'} \rho(r'')r^{\prime\prime 2} dr''.
\ee
A change of integration variable to $\xi=1/r'$, combined with a
division of the inner integral at $r''=r\leq 1/\xi$, leaves
\be
V(r)=-4\pi Gm\int_0^{1/r} d\xi\left[\int_0^r \rho(r'')r^{\prime\prime 2} dr''
                                  +\int_r^{1/\xi}  \rho(r'')r^{\prime\prime 2} dr''\right].
\ee
By changing the order of integration, we have
\be
V(r)=-4\pi Gm\left[\int_0^r \rho(r'') r^{\prime\prime 2} dr''\int_0^{1/r}d\xi
                    +\int_r^\infty \rho(r'') r^{\prime\prime 2} dr''\int_0^{1/r''} d\xi\right].
\ee
The $\xi$ integrals are now trivial.
Use of $\rho=\frac{m}{4\pi r^2}|u(r)|^2$ reduces the expression for $V$ to
\be \label{eq:NRpotl}
V(r)=-Gm^2\left[ \int_0^r\frac{|u(r'')|^2}{r} dr'' + \int_r^\infty \frac{|u(r'')|^2}{r''} dr''\right].
\ee
This makes the Schr\"odinger equation (\ref{eq:SchrEqn}) a nonlinear equation.

We solve the combination of (\ref{eq:SchrEqn}) and (\ref{eq:NRpotl}) self consistently
by iteration.  This is done numerically, with a cutoff in radius $r_{\rm max}$
taken large enough to not influence the solution significantly
and with the second term in (\ref{eq:NRpotl}) computed as an
integral from zero to the cutoff minus the integral from zero to $r$:
\be
\int_r^{r_{\rm max}} \frac{|u(r'')|^2}{r''} dr''
   =\int_0^{r_{\rm max}} \frac{|u(r'')|^2}{r''} dr''-\int_0^r \frac{|u(r'')|^2}{r''} dr''.
\ee
The wave function $u(r'')$ goes to zero rapidly enough at 
$r''=0$ to avoid a singularity.  The integrals are approximated
by the trapezoidal rule, which generates values for the potential
$V$ on the chosen grid, and the Schr\"odinger equation is discretized
on the same grid, to produce a matrix eigenvalue problem from the
finite-difference representation.  

For the numerical calculation, we introduce dimensionless forms of
Eqs.~(\ref{eq:SchrEqn}) and (\ref{eq:NRpotl}).  The natural length scale is the
gravitational Bohr radius $a=1/Gm^3$, and the natural energy
scale is $G^2m^5$.  In terms of these we have a dimensionless
energy $\epsilon\equiv E/G^2m^5$ and dimensionless radial
coordinate $\zeta\equiv r/a$ and define a dimensionless
wave function $\tilde{u}\equiv \sqrt{a}u$ and potential
$\tilde{V}\equiv V/G^2m^5$. The Schr\"odinger--Newton system becomes
\be
-\frac{1}{2}\frac{d^2\tilde{u}}{d\zeta^2}+\tilde{V}(\zeta)\tilde{u}(\zeta)=\epsilon\tilde{u}(\zeta)
\ee
and 
\be
\tilde{V}(\zeta)=-\frac{1}{\zeta}\int_0^\zeta |\tilde{u}(\zeta')|^2 d\zeta' 
       - \int_0^{\zeta_{\rm max}} \frac{|\tilde{u}(\zeta')|^2}{\zeta'} d\zeta'.
       +\int_0^\zeta \frac{|\tilde{u}(\zeta')|^2}{\zeta'} d\zeta'.
\ee

\begin{figure}[ht]
\vspace{0.2in}
\includegraphics[width=15cm]{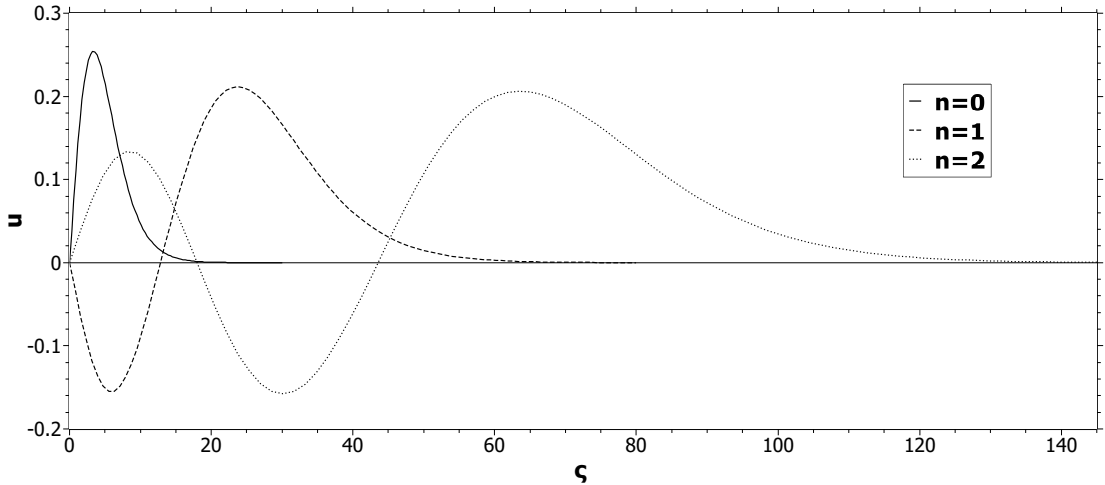}
\caption{\label{fig:u-nonrel}
Plots of the modified radial wave function $u$ 
in units of $\sqrt{a}$, with $a$ the
gravitational Bohr radius $1/Gm^3$, for the ground-state
soliton ($n=0$) and two excited cases ($n=1,2$), all for
the nonrelativistic Schr\"odinger--Newton problem.  
Here $n$ is the number of radial nodes.
The associated energies are listed in the first row of
Table~\protect\ref{tab:eigenvalues}.  The dimensionless
radial coordinate $\zeta$ is rescaled by $a$.
}
\end{figure}

\begin{figure}[ht]
\vspace{0.2in}
\includegraphics[width=15cm]{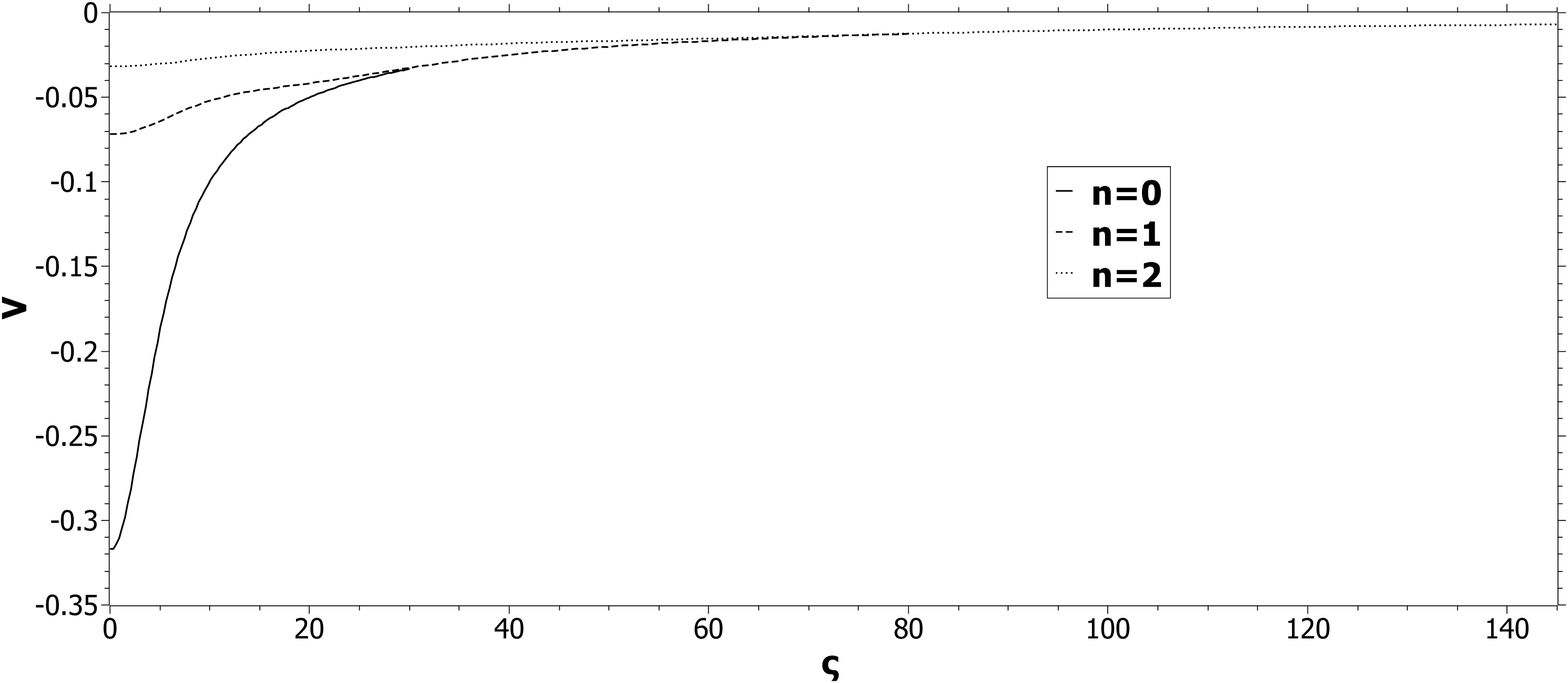}
\caption{\label{fig:V-nonrel}
Same as Fig.~\protect\ref{fig:u-nonrel} but for
the gravitational potential $V$, in units of $G^2m^5$.
Each radial eigenfunction, labeled by $n$,
has its own self-consistent potential.
}
\end{figure}

We compute not only the lowest state but
also radial excitations, for which the potential is again obtained
self consistently and therefore different for each 
state.\footnote{For perturbations of these solitons,
see \protect\cite{Zagorac}.}  Our results are consistent with earlier 
calculations~\cite{Moroz,Bernstein,Tod,HarrisonMorozTod,Harrison}.  
Table~\ref{tab:eigenvalues} lists the results for the ground state 
and the lowest two radial excitations.  Figures~\ref{fig:u-nonrel} and
\ref{fig:V-nonrel} show the
modified radial wave function $u$ and the shape of
the gravitational potential $V$ for these same cases.
However, our main purpose is to compare with a fully
relativistic calculation, which we consider in the
next section.

\begin{table}[ht]
\caption{\label{tab:eigenvalues}
Energy eigenvalues in units of $G^2m^5$ for the nonrelativistic (NR) Schr\"odinger--Newton
solitons and the relativistic Einstein--Klein--Gordon solitons, the latter
being associated with various values of the rescaled Schwarzschild
radius $\zeta_S=2G^2m^4$.  Here $G$ is Newton's gravitational constant and
$m$ is the mass associated with the self-gravitating field.
For the relativistic results, `PF' indicates the perfect-fluid model
and `GR', the direct general relativistic calculation.
Relativistic effects increase with $\zeta_S$.
Two radial excitations are listed, with $n$ the number of
radial nodes in the wave function.
}
\begin{center}
\begin{tabular}{l|l|rrr}
\hline \hline
$\zeta_S$ & type & $n=0$ & $n=1$ & $n=2$ \\
\hline
 --     & NR & -0.1628 & -0.0309 & -0.0125 \\
0.01  & PF & -0.1631 & -0.0308 & -0.0125 \\
      & GR & -0.1631 & -0.0308 & -0.0125 \\
0.1   & PF & -0.1663 & -0.0308 & -0.0126 \\
      & GR & -0.1657 & -0.0309 & -0.0125 \\
0.2   & PF & -0.1701 & -0.0311 & -0.0126 \\
      & GR & -0.1688 & -0.0310 & -0.0126 \\
0.5   & PF & -0.1839 & -0.0315 & -0.0126 \\
      & GR & -0.1795 & -0.0313 & -0.0126 \\
1.0   & PF & -0.2218 & -0.0322 & -0.0127 \\
      & GR & -0.2045 & -0.0318 & -0.0127 \\
\hline \hline
\end{tabular}
\end{center}
\end{table}

\section{Einstein--Klein--Gordon solitons}  \label{sec:KleinGordon}

\subsection{Klein--Gordon equation in curved spacetime}  \label{sec:KleinGordonEqn}

For a proper representation of gravity in a relativistic
formulation, we must of course invoke spacetime curvature
as represented by a metric $g_{\mu\nu}$.  We are interested
in static, spherically symmetric solitons, which means that
the metric must have this symmetry.  The KG equation
for a scalar particle of mass $m$ in this curved spacetime is
\be
\frac{1}{\sqrt{-g}}\partial_\mu\left[\sqrt{-g}g^{\mu\nu}\partial_\nu\Psi(x)\right]
  +m^2\Psi=0.
\ee
We choose spherical coordinates such that $g_{\mu\nu}$ is diagonal
and the line element is
\be
ds^2=g_{00}dt^2-g_{rr}dr^2-r^2d\theta^2-r^2\sin^2\theta d\phi^2,
\ee
for which $g\equiv{\rm det}[g_{\mu\nu}]=-g_{00}g_{rr} r^4\sin^2\theta$,
with $g_{00}$ and $g_{rr}$ functions only of the radial coordinate $r$.
The KG equation then takes the form
\bea
\frac{1}{\tilde{g}r^2\sin\theta}
  \left[\frac{\partial}{\partial t}\left(\frac{\tilde{g}r^2\sin\theta}{g_{00}}\frac{\partial \Psi}{\partial t}\right)\right.
        &-&\frac{\partial}{\partial r}\left(\frac{\tilde{g}r^2\sin\theta}{g_{rr}}\frac{\partial\Psi}{\partial r}\right)
        -\frac{\partial}{\partial\theta}\left(\tilde{g}\sin\theta\frac{\partial\Psi}{\partial\theta}\right) \\
        &&-\left.\frac{\partial}{\partial\phi}\left(\frac{\tilde{g}}{\sin\theta}\frac{\partial\Psi}{\partial\phi}\right)\right]
          +m^2\Psi=0,  \nonumber
\eea
where\footnote{For the Schwarzschild metric, $\tilde{g}=1$. In \protect\cite{Lehn} this
was incorrectly assumed true for other computed metrics, making
any non-Schwarzschild results there only qualitative.}
$\tilde{g}\equiv\sqrt{g_{00}g_{rr}}$ so that 
$\sqrt{-g}=\tilde{g}r^2\sin\theta$.  For a static metric and $\tilde{g}$ 
independent of angles, this reduces to 
\be
\frac{1}{g_{00}}\frac{\partial^2\Psi}{\partial t^2}
   -\frac{1}{\tilde{g}r^2}\frac{\partial}{\partial r}\left(\frac{\tilde{g}r^2}{g_{rr}}\frac{\partial\Psi}{\partial r}\right)
   +\frac{L^2}{r^2}\Psi+m^2\Psi=0,
\ee
with the usual definition of $L^2$ as
\be
L^2\equiv-\left[\frac{1}{\sin\theta}\frac{\partial}{\partial\theta}\left(\sin\theta\frac{\partial}{\partial\theta}\right)
        +\frac{1}{\sin^2\theta}\frac{\partial^2}{\partial\phi^2}\right].
\ee

We then apply separation of variables with $\Psi=\tau(t)R_l(r)Y_{lm}(\theta,\phi)$
and isolate the $t$ and $r$ dependence as
\be
\frac{1}{\tau}\frac{d^2\tau}{dt^2}
  =\frac{g_{00}}{\tilde{g}r^2R_l}\frac{d}{dr}\left(\frac{\tilde{g}r^2}{g_{rr}}\frac{dR_l}{dr}\right)
     -\left[\frac{l(l+1)}{r^2}+m^2\right]g_{00}\equiv -E^2.
\ee
Here $-E^2$ is the separation constant with $E$ obviously interpreted
as an energy and $E-m$ the binding energy.  The time-dependent $\tau$ function is
just $e^{\pm iEt}$. 

We focus on the radial equation:
\be  \label{eq:radial}
-\frac{1}{\tilde{g}r^2}\frac{d}{dr}\left(\frac{\tilde{g}r^2}{g_{rr}}\frac{dR_l}{dr}\right)
     +\frac{l(l+1)}{r^2}R_l+m^2R_l= \frac{E^2}{g_{00}}R_l.
\ee
To facilitate the numerical solution of this equation, we wish to eliminate any first-derivative
terms; a finite-difference approximation will then yield a symmetric matrix representation.
To accomplish this, we introduce a modified radial
wave function $u_l(r)\equiv h(r)R_l(r)$ with $h(r)$ chosen to eliminate any 
first-derivative terms in
\be
\frac{d}{dr}\left(\frac{\tilde{g}r^2}{g_{rr}}\frac{dR_l}{dr}\right)
= \frac{d}{dr}\left(\frac{\tilde{g}r^2}{g_{rr}h}\right)\left(\frac{du_l}{dr}-\frac{h'}{h}u_l\right)
  +\frac{\tilde{g}r^2}{g_{rr}h}\left(\frac{d^2u_l}{dr^2}-\frac{h'}{h}\frac{du_l}{dr}
         +\frac{(h')^2}{h^2}u_l-\frac{h''}{h}u_l\right).
\ee
The coefficient of $\frac{du_l}{dr}$ is set to zero:
\be   \label{eq:h-condition}
\frac{d}{dr}\left(\frac{\tilde{g}r^2}{g_{rr}h}\right)
  -\frac{\tilde{g}r^2}{g_{rr}h}\frac{h'}{h}=0.
\ee
Except for a multiplicative constant, the solution is 
\be
h=r\sqrt{\frac{\tilde{g}}{g_{rr}}}.
\ee
The constant in $h$ is irrelevant, given that $h$ appears
only in ratios, or can be viewed as absorbed into
the normalization of $u_l$.  The condition (\ref{eq:h-condition}) on $h$
also eliminates two terms proportional to $u_l$,
leaving
\be
\frac{d}{dr}\left(\frac{\tilde{g}r^2}{g_{rr}}\frac{dR_l}{dr}\right)
= \frac{\tilde{g}r^2}{g_{rr}h}\left(\frac{d^2u_l}{dr^2}-\frac{h''}{h}u_l\right).
\ee
This provides a relatively simple equation for $u_l$:
\be
-\frac{d^2u_l}{dr^2}+\frac{h''}{h}u_l+\left[\frac{l(l+1)}{r^2}+m^2\right]g_{rr}u_l
    =\frac{g_{rr}}{g_{00}}E^2u_l.
\ee
Solutions of this and the original radial equation for a
fixed metric, particularly the Schwarzschild metric, have
been considered numerically~\cite{Lehn} and 
analytically~\cite{Smith,Rowan,Elizalde,Qin,Li}.

The normalization condition is
\be
1=\int_0^\infty |R_l|^2\sqrt{g_{rr}}r^2\,dr=\int_0^\infty |u_l|^2 \frac{g_{rr}}{\sqrt{g_{00}}} dr.
\ee
The probability density is 
\be
\rho_{lm}=|R_l|^2 |Y_{lm}|^2=\frac{|u_l|^2}{h^2}|Y_{lm}|^2.
\ee
%

\subsection{Perfect-fluid approximation}  \label{sec:PF}

To generate a spherically symmetric metric from a mass density $m\rho_{lm}$,
we consider only $l=0$ and define the mass density as
\be
\rho(r)=\frac{m|u_o|^2}{4\pi h^2}.
\ee
This mass density is the source for the computation of the metric.  When
viewed as a self-consistent solution in a Hartree approximation to
a many-body bosonic state, this density can be modeled as a perfect
fluid in hydrostatic equilibrium.  The metric is then determined
by the TOV equation~\cite{Tolman,Oppenheimer,Carroll,ExactSolns}
for the pressure $p(r)$
\be  \label{eq:TOV}
\frac{dp}{dr}=-G\frac{[\rho(r)+p(r)][\mu(r)+4\pi r^3 p(r)]}{r[r-2G\mu(r)]},
\ee
with the mass function
\be  \label{eq:massfunction}
\mu(r)\equiv4\pi\int_0^r \rho(r')r^{\prime 2}dr'.
\ee
For the spherically symmetric static case, the GR equations are
then satisfied by solutions of the form~\cite{ExactSolns}
\be
g_{00}=e^{2A(r)}, \;\; g_{rr}=\frac{1}{1-2G\mu(r)/r},
\ee
with the metric function $A(r)$ determined by
\be \label{eq:Phi}
\frac{dA}{dr}=G\frac{\mu(r)+4\pi r^3 p(r)}{r[r-2G\mu(r)]}.
\ee

These three equations, (\ref{eq:TOV}), (\ref{eq:massfunction}), and
(\ref{eq:Phi}), form a coupled set of integro-differential equations 
for the metric components with
the boundary conditions $\mu(0)=0$, $A(r)\sim\ln\sqrt{1-2G\mu(r)/r}$,
$p(\infty)=0$.  The form of $A$ applies for $r$ large enough that
$\rho$ is effectively zero and all of the mass is contained.
The mass function $\mu$ does not reach $m$ even at this range
because the fluid structure implicitly assumes internal
gravitational binding energy.  Thus $\mu(\infty)$ is equal to
the mass $m$ minus the gravitational binding energy of the fluid,
and $\mu$ is computed without a
curvature contribution to the Jacobian~\cite{Carroll}.

Just as for the nonrelativistic case, there is a natural 
length scale, the gravitational Bohr radius $a=1/Gm^3$,
and an energy scale $G^2m^5$.  From the latter 
we define the dimensionless energy parameter $\epsilon$
in terms of the binding energy
\be
\Delta E=E-m=\frac{Gm^2}{a}\epsilon=G^2m^5\epsilon.
\ee
Unlike the nonrelativistic case, there is another
length scale, the Schwarzschild radius $r_S=2Gm$.
Therefore, in addition to the rescaled
radial coordinate $\zeta\equiv r/a$, we define
a dimensionless Schwarzschild radius $\zeta_S=r_S/a=2G^2m^4$.  
As shown in \cite{Lehn} and reproduced in Appendix~\ref{sec:NonrelLimit}, the 
magnitude of $\zeta_S$ determines the importance of
relativistic effects.

These parameters can be used to rescale the coupled 
system of equations, including the reduced KG
equation, which must be solved self consistently.
We define
\be
\tilde{u}_l=\sqrt{a}u_l,\;\;
\tilde{h}=h/a, \;\;
\tilde\rho=\frac{4\pi a^3}{3m}\rho=\frac{|\tilde{u}_0|^2}{3\tilde{h}^2},
\ee
and
\be
\tilde\mu(\zeta)=\frac{\mu(a\zeta)}{m}=\int_0^\zeta |\tilde{u}_0|^2\sqrt{\frac{g_{rr}}{g_{00}}}d\zeta', \;\;
\tilde{p}(\zeta)=\frac{4\pi a^3}{3m}p(a\zeta).
\ee
The function $A$ is already dimensionless.  The full coupled system
of equations becomes\footnote{In Eq.~(16) of \cite{Lehn}, which
is the equivalent of the first equation here for general $l$, there is
an $m^2$ factor that should be $2/\zeta_S$ instead.  Also, the terms on the right are
slightly different because here the energy scale is larger by a factor
of 2, to be consistent with earlier work on the Schr\"odinger--Newton
problem~\protect\cite{Harrison}.}
\be  \label{eq:KGscaled}
-\frac{d^2\tilde{u}_0}{d\zeta^2}+\left[\frac{\tilde{h}''}{\tilde{h}}
                 +\frac{2}{\zeta_S}g_{rr}\left(1-\frac{1}{g_{00}}\right)\right]\tilde{u}_0
                 =\left(2\epsilon+\zeta_S\frac{\epsilon^2}{2}\right)\frac{g_{rr}}{g_{00}}\tilde{u}_0,
\ee
\be \label{eq:muscaled}
\frac{d\tilde\mu}{d\zeta}=|\tilde{u}_0|^2\sqrt{\frac{g_{rr}}{g_{00}}}, \;\; \tilde\mu(0)=0,
\ee
\be \label{eq:pressurescaled}
\frac{d\tilde{p}}{d\zeta}=-\frac{\zeta_S}{2}\frac{[\tilde\rho(\zeta)+\tilde{p}(\zeta)][\tilde\mu(\zeta)+3\zeta^3\tilde{p}(\zeta)]}
                                                 {\zeta[\zeta-\zeta_S\tilde\mu(\zeta)]}, \;\;
\tilde{p}(\infty)=0,
\ee
and
\be \label{eq:Phiscaled}
\frac{dA}{d\zeta}=\frac{\zeta_S}{2}\frac{\tilde\mu(\zeta)+3\zeta^3\tilde{p}(\zeta)}{\zeta[\zeta-\zeta_S\tilde\mu(\zeta)]}, \;\;
A\sim\ln\sqrt{1-\zeta_S\tilde\mu(\zeta)/\zeta},
\ee
with
\be \label{eq:metricscaled}
g_{00}=e^{2A},\;\;
g_{rr}=\frac{1}{1-\zeta_s\tilde\mu(\zeta)/\zeta},
\ee
and the double prime in $\tilde{h}''$ meaning $d^2\tilde{h}/d\zeta^2$.

This system is solved self consistently starting from an initial guess for
the metric, taken as flat inside $\zeta_S$ and the Schwarzschild metric
with radius $\zeta_S$ for $\zeta>\zeta_S$. The KG equation
(\ref{eq:KGscaled}) is solved for $\tilde{u}_0$ and
$\epsilon$.  This determines a guess for the density for which
the mass and pressure functions are computed by outward and inward
integration of (\ref{eq:muscaled}) and (\ref{eq:pressurescaled}),
respectively.  Finally, (\ref{eq:Phiscaled}) can be integrated inward
to find $A$.  The expressions in (\ref{eq:metricscaled}) can
then be evaluated to determine an improved metric.  The cycle
begins again and iterates until convergence to an appropriate
tolerance.  Some further details of the numerical calculation
are given in Appendix~\ref{sec:numerics}.

\begin{figure}[hb]
\vspace{0.2in}
\includegraphics[width=15cm]{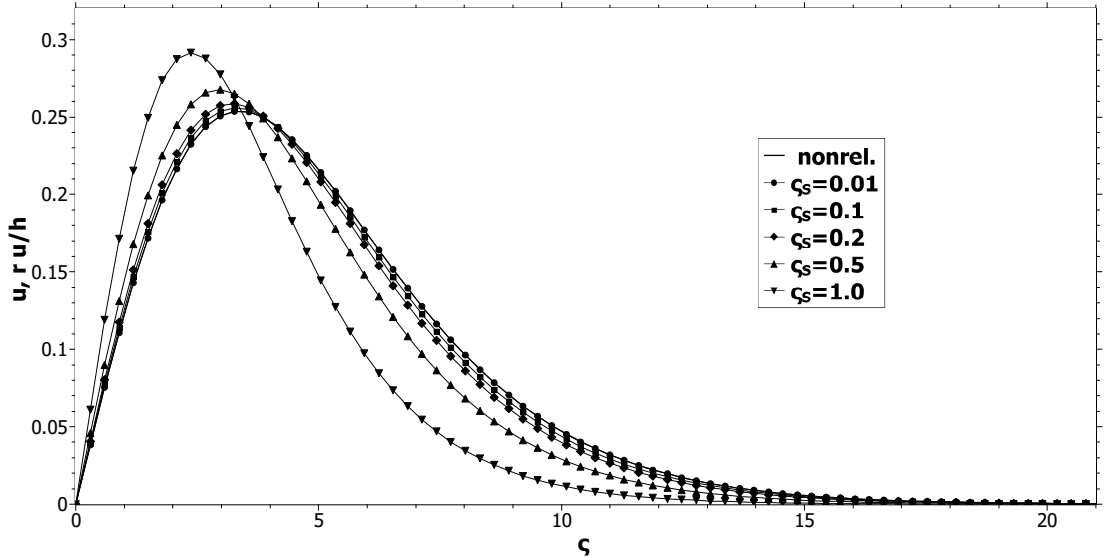}
\caption{\label{fig:n0amplitudes}
Plots of the ground-state radial probability amplitudes for the 
nonrelativistic and relativistic perfect-fluid cases
as functions of the dimensionless radial coordinate $\zeta$.
The relativistic amplitudes are distinguished by the different
values of the rescaled Schwarzschild radius $\zeta_S$.
For the Schr\"odinger--Newton
soliton, the amplitude is simply the modified radial
wave function $\sqrt{a}u$; for the Einstein--Klein--Gordon solitons,
the amplitude is $\sqrt{a}u_0\sqrt{g_{rr}/\sqrt{g_{00}}}$.
The KG amplitudes are made comparable by
a rescaling to match the Schr\"odinger--Newton peak height
to that of the amplitude for $\zeta_S=0.01$, which is
then indistinguishable.
The associated energies are listed in the various rows of
Table~\protect\ref{tab:eigenvalues}. 
}
\end{figure}

The results for eigenenergies are listed in Table~\ref{tab:eigenvalues},
with the type designated as `PF'.
As $\zeta_S$ is increased, the states become more deeply bound,
particularly for the ground state.  This is consistent with
the change in the probability amplitudes, plotted in 
Figs.~\ref{fig:n0amplitudes}, where
the peaks are shifted toward $\zeta=0$ as $\zeta_S$ is
increased. It is also consistent with the analysis 
by Brizuela and Duran-Cabac\'es~\cite{Brizuela}
of relativistic corrections to the nonrelativistic case,
showing that the self-gravitation is increased.
For small $\zeta_S$, the amplitudes agree with
the nonrelativistic amplitudes, the two being
indistinguishable in the plot.  For radial excitations
the relativistic effects are far less, and
the amplitudes all match the nonrelativistic shape
for the full range of $\zeta_S$ values considered.

\subsection{Direct general relativistic calculation}  \label{sec:GR}

The GR field equations $G_{\mu\nu}=8\pi GT_{\mu\nu}$
can be solved directly in this static, spherically symmetric case.
The stress-energy tensor for the scalar field is~\cite{Giulini}
\be
T_{\mu\nu}=\frac{1}{2m}\left[(\partial_\mu\Psi)(\partial_\nu\Psi^*)
                      +(\partial_\mu\Psi^*)(\partial_\nu\Psi)
                      -g_{\mu\nu}(\partial^\lambda\Psi)(\partial_\lambda\Psi^*)\right]
            -g_{\mu\nu}\frac{m^2}{2}|\Psi|^2.
\ee
Given this as the source, with $\Psi=R(r)e^{\pm iEt}/\sqrt{4\pi}$, 
and the metric coefficients written as 
\be
g_{00}=e^{2A(r)}\;\;\mbox{and}\;\;g_{rr}=e^{2B(r)},
\ee
the field equations become~\cite{Giulini}
\be
e^{2A}\left[\frac{1}{r^2}-e^{-2B}\left(\frac{1}{r^2}-\frac{2}{r}B'\right)\right]
=2 G\left[e^{2A}|R|^2+\frac{1}{2m}e^{2(A-B)}|R'|^2+\frac{E^2}{2m}|R|^2\right],
\ee
\be
\frac{1}{r^2}(1-e^{2B})+\frac{2}{r}A'
=2G\left[-\frac{m}{2}e^{2B}|R|^2+\frac{1}{2m}|R'|^2+\frac{E^2}{2m}e^{2(B-A)}|R|^2\right],
\ee
\bea  \label{eq:GR3}
r^2 e^{-2B}\left[(A')^2-A'B'+A''+\frac{A'-B'}{r}\right] 
&=&2G\left[-\frac{m}{2}r^2|R|^2-\frac{r^2}{2m}e^{-2B}|R'|^2  \right. \\
&& \rule{1in}{0pt} \left.+\frac{r^2E^2}{2m}e^{-2A}|R|^2\right].  \nonumber
\eea
Here a prime indicates differentiation with respect to $r$.  The KG equation
for $R$, Eq.~(\ref{eq:radial}), can be written in terms of the same metric functions as
\be
-R''-\left[\frac{2}{r}+A'-B'\right]R'+m^2 e^{2B}R=E^2e^{2(B-A)}R.
\ee
The third GR equation, Eq.~(\ref{eq:GR3}), can be derived
from this radial KG equation and the first two GR equations.

The dimensionless forms of the first two GR equations are
\bea
\frac{dA}{d\zeta}&=&\frac{1}{2\zeta}(e^{2B}-1)-\frac{\zeta_S\zeta}{4}e^{2B}\tilde{R}^2
  +\frac{\zeta_S^2\zeta}{8}\left(\frac{d\tilde{R}}{d\zeta}\right)^2
  +\frac{\zeta_S\zeta}{4}\left[1+\frac12\zeta_S\epsilon\right]^2e^{2(B-A)}\tilde{R}^2, \\
\frac{dB}{d\zeta}&=&-\frac{1}{2\zeta}(e^{2B}-1)+\frac{\zeta_S\zeta}{4}e^{2B}\tilde{R}^2
  +\frac{\zeta_S^2\zeta}{8}\left(\frac{d\tilde{R}}{d\zeta}\right)^2
  +\frac{\zeta_S\zeta}{4}\left[1+\frac12\zeta_S\epsilon\right]^2e^{2(B-A)}\tilde{R}^2,
\eea
with $\tilde{R}=a^{3/2}R=\tilde{u}/\tilde{h}$.  Given the
radial function $\tilde{u}$, these equations can be 
solved numerically for the metric functions $A$ and $B$,
and the radial KG equation is solved self consistently.

\begin{figure}[hb]
\vspace{0.2in}
\includegraphics[width=15cm]{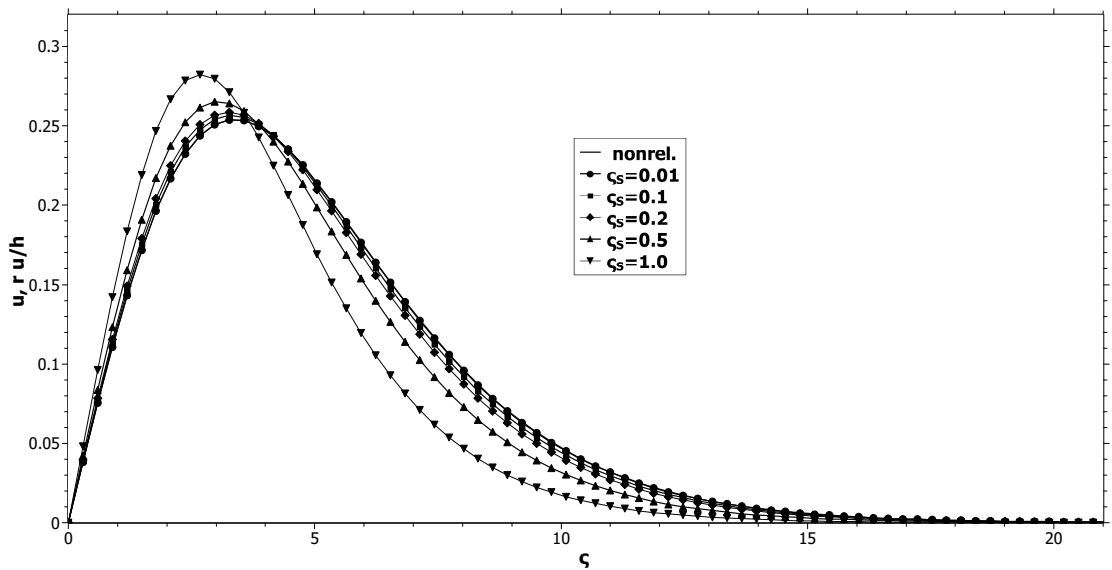}
\caption{\label{fig:GRamplitudes}
Same as Fig.~\ref{fig:n0amplitudes} but for the direct
general relativistic calculation.
}
\end{figure}

\begin{figure}[hb]
\vspace{0.2in}
\includegraphics[width=15cm]{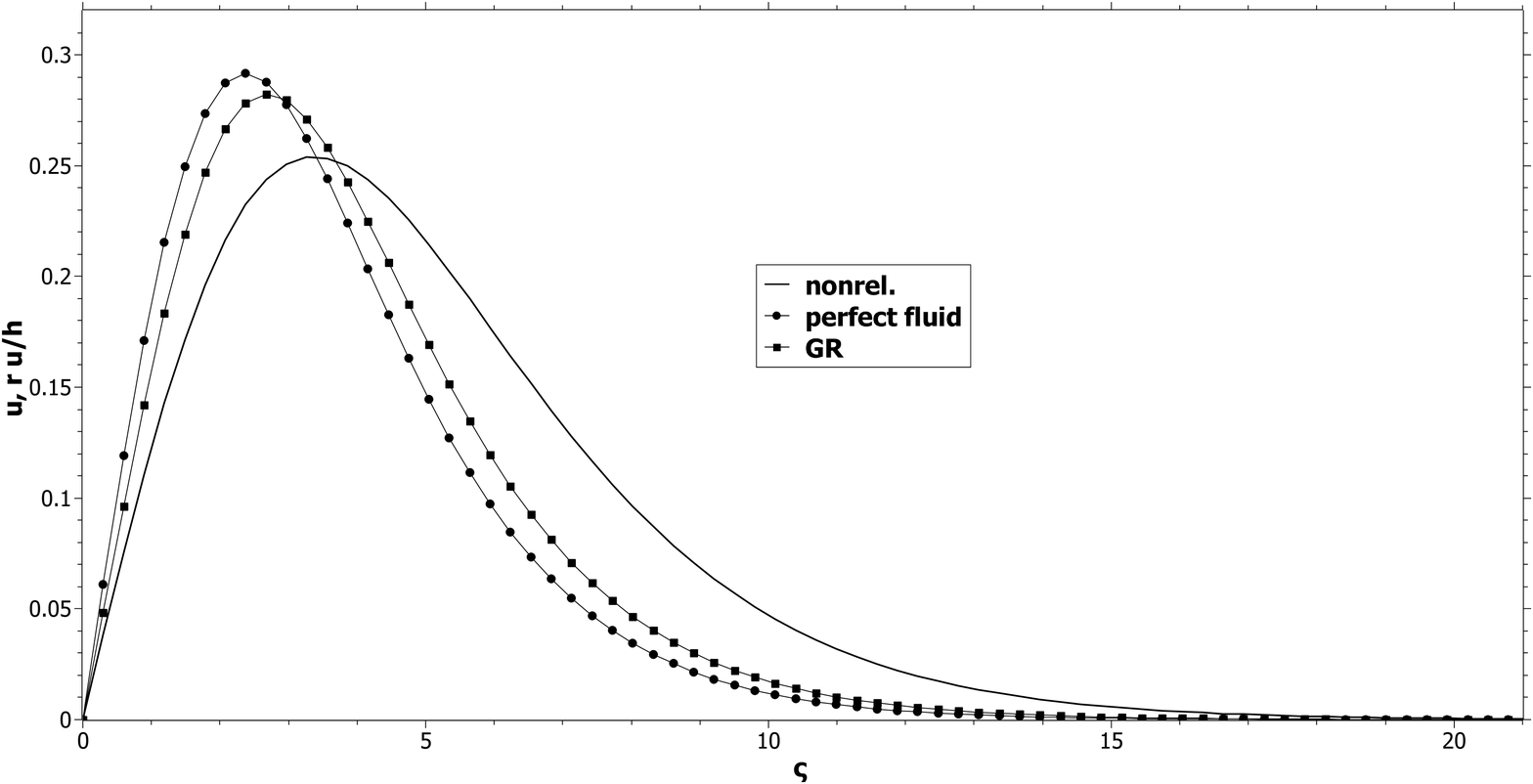}
\caption{\label{fig:compare}
Comparison of the wave functions for the perfect-fluid
model and the direct general relativistic calculation,
for $\zeta_S=1$, and the nonrelativistic wave function.
}
\end{figure}

The result for the energy eigenvalues $\epsilon$ are listed
in Table~\ref{tab:eigenvalues}, designated as the type `GR'.  
Except for the largest value
of $\zeta_S$, they are not significantly different from those
of the perfect-fluid model.  The wave functions are plotted
in Fig.~\ref{fig:GRamplitudes} and, for $\zeta_S=1$, compared
with the wave function in the perfect-fluid model in
Fig.~\ref{fig:compare}.  Again, the ground state in
the relativistic case is more deeply bound, but not as
much as in the perfect-fluid model; this can be seen
explicitly in Fig.~\ref{fig:compare} and Table~\ref{tab:eigenvalues}.

\section{Summary}  \label{sec:summary}

We have shown that the relativistic version of the
Schr\"odinger--Newton problem for scalar particles
can be solved for Einstein--Klein--Gordon solitons
in spherically symmetric spacetimes.  This includes
radial excitations. We consider both a perfect-fluid model, 
consistent with a Hartree approximation to a bosonic star,
and the fundamental GR equations with the stress-tensor
of the KG field. The results for the Schr\"odinger--Newton problem
are recovered in the nonrelativistic limit, which
is controlled by the ratio of the Schwarzschild radius
to the gravitational Bohr radius for the given mass.

The eigenenergies obtained are listed in Table~\ref{tab:eigenvalues}.
The relativistic cases are more deeply bound than the
nonrelativistic case, particularly for the ground state.
This can also be seen in the amplitudes, as plotted in
Figs.~\ref{fig:n0amplitudes}, \ref{fig:GRamplitudes}, and
\ref{fig:compare}, where the relativistic peaks occur
at smaller radii.  This is consistent with the
findings of Brizuela and Duran-Cabac\'es~\cite{Brizuela}
in their analysis of relativistic corrections to
the nonrelativistic Schr\"odinger--Newton problem.
We also find that the perfect-fluid model binds more
deeply than occurs for the fundamental GR equations
that use only the stress-energy tensor of the scalar
field.  Apparently, the additional assumption of
hydrostatic equilibrium increases the energy
density and consequently the spatial curvature.

The restriction to spherical symmetry
can be relaxed to consider cylindrical symmetry.
This has been done at least partially
for the Schr\"odinger--Newton problem~\cite{Harrison},
though with an unnecessary assumption of
a cylindrically symmetric wave function with $L_z=0$.
A more complete nonrelativistic calculation could
be done as well as consideration of a
relativistic formulation~\cite{Bronnikov}.

Our approach represents a form of semi-classical gravity
where the matter fields are treated quantum
mechanically but gravity classically.  It requires
self-consistent solutions for the metric and
the quantum particle amplitude.  The results of
such computations may provide a check on the 
structure of a theory of quantum gravity.  

\acknowledgments
This work was supported in part by 
the Minnesota Supercomputing Institute 
and the Research Computing and Data Services
at the University of Idaho through
grants of computing time.

\appendix

\section{Nonrelativistic limit}  \label{sec:NonrelLimit}

For completeness, we repeat the argument from \cite{Lehn} that
$\zeta_S$ controls the importance of relativistic effects.
For simplicity, we consider the Schwarzschild geometry,
for which $g_{00}=1-2Gm/r=1-\zeta_S/\zeta=1/g_{rr}$.
In this case, we have $\tilde{g}=\sqrt{g_{00} g_{rr}}=1$
and $h=r/\sqrt{g_{rr}}=a\zeta/\sqrt{g_{rr}}$.
The $\tilde{h}''/\tilde{h}$ term in the modified
KG equation (\ref{eq:KGscaled}) is then
\be
\frac{\tilde{h}''}{\tilde{h}}=-\frac{\zeta_S^2}{4\zeta^2(\zeta^2-\zeta_S^2)},
\ee
and the modified radial equation becomes
\be
-\frac{d^2\tilde{u}_0}{d\zeta^2}-\frac14\frac{\zeta_S^2}{\zeta^2(\zeta-\zeta_S)^2}\tilde{u}_0
                           -\frac{2}{\zeta}\frac{1}{(1-\zeta_S/\zeta)^2}\tilde{u}_0
                =\frac{2\epsilon+\frac12\zeta_S\epsilon^2}{(1-\zeta_S/\zeta)^2}\tilde{u}_0.
\ee
Keeping $\zeta_S$ to first order, we obtain
\be
-\frac{d^2\tilde{u}_0}{d\zeta^2}
                -\frac{2}{\zeta}(1+2\zeta_S/\zeta)\tilde{u}_0
                =(2\epsilon(1+2\zeta_S/\zeta)+\frac12\zeta_S\epsilon^2)\tilde{u}_0,
\ee
which can be rearranged as
\be
-\frac{d^2\tilde{u}_0}{d\zeta^2}-\frac{2}{\zeta}\tilde{u}_0
-\zeta_S\left[\frac{4}{\zeta^2}
-\frac{4\epsilon}{\zeta}+\frac12\epsilon^2\right]\tilde{u}_0
                =2\epsilon\tilde{u}_0.
\ee
The $\zeta_S$ terms are then revealed to be corrections to the ordinary Coulomb problem
of Newtonian gravity.

\section{Details of the numerical calculation}  \label{sec:numerics}

Just as for the nonrelativistic case, the infinite range of the
radial coordinate is truncated at a distant point and the 
density and pressure are assumed to be zero beyond that point.
The scaled KG equation is represented by a matrix
equation obtained from finite-difference approximations to
the derivatives of $\tilde{u}_0$ and $\tilde{h}$
on an equally spaced grid.  The metric is then
computed on this grid by solving the first-order equations
for $\tilde\mu$, $\tilde{p}$ and $A$ in the perfect-fluid model,
or for $A$ and $B$ in the fundamental GR equations,
with a second-order 
Runge--Kutta method, utilizing a matching step size.  This choice 
has an error term consistent with
the chosen finite-difference approximation to the KG
equation.  For better accuracy, one could of course use
higher order methods, but these were sufficient for the
purpose of comparing the nonrelativistic and relativistic
results, with approximately four significant figures in
the values of scaled energies.

In order that the matrix representation of the KG equation be symmetric,
we introduce a new function $\bar{u}\equiv\tilde{u}\sqrt{g_{rr}/g_{00}}$
and multiply (\ref{eq:KGscaled}) by $\sqrt{g_{rr}/g_{00}}$.
We also define $\lambda\equiv 2\epsilon+\zeta_S\epsilon^2/2$
as the direct eigenvalue of the matrix.  The differential
equation solved numerically to obtain the KG eigenstates is
actually
\be
-\sqrt{\frac{g_{00}}{g_{rr}}}\frac{d^2}{d\zeta^2}\left[\sqrt{\frac{g_{00}}{g_{rr}}}\bar{u}\right]
  +\frac{g_{00}}{g_{rr}}\frac{\tilde{h}''}{\tilde{h}}\bar{u}
  +\frac{2}{\zeta_S}g_{00}\left[1-\frac{1}{g_{00}}\right]\bar{u}=\lambda\bar{u}.
\ee
The third term on the left-hand side is best evaluated as $2e^A\sinh A$,
which comes from $g_{00}=e^{2A}$,
rather than in its explicit form, because $g_{00}$ can be close to 1.
Also, the scaled binding energy is best extracted from $\lambda$
by a rearrangement of the quadratic formula
\be
\epsilon=\frac{\lambda}{1+\sqrt{1+\zeta_s\lambda/2}}.
\ee
%


\end{document}